\documentstyle[aps,pra,eqsecnum,preprint,tighten,psfig,floats]{revtex}
\newcommand{\openeen}{\leavevmode\hbox{\small1\kern-3.8pt\normalsize1}}
\newcommand{\gtequiv}{\lower2pt\hbox{$\:\stackrel{>}{
\scriptstyle\sim}\:$}}
\newcommand{\ltequiv}{\lower2pt\hbox{$\:\stackrel{<}{
\scriptstyle\sim}\:$}}
\begin{document}
\draft
\title{PHOTON STATISTICS OF A RANDOM LASER}
\author{C.W.J. BEENAKKER}
\address{Instituut-Lorentz, Leiden University\\
P.O. Box 9506, 2300 RA Leiden, The Netherlands}
\date{May 1998}
\maketitle
\begin{abstract}
A general relationship is presented between the statistics of thermal radiation
from a random medium and its scattering matrix $S$. Familiar results for
black-body radiation are recovered in the limit $S\rightarrow 0$. The mean
photo\-count $\bar{n}$ is proportional to the trace of $\openeen -S\cdot
S^{\dagger}$, in accordance with Kirchhoff's law relating emissivity and
absorptivity. Higher moments of the photo\-count distribution are related to
traces of powers of $\openeen -S\cdot S^{\dagger}$, a generalization of
Kirchhoff's law. The theory can be applied to a random amplifying medium (or
``random laser'') below the laser threshold, by evaluating the Bose-Einstein
function at a negative temperature. Anomalously large fluctuations are
predicted in the photo\-count upon approaching the laser threshold, as a
consequence of overlapping cavity modes with a broad distribution of spectral
widths.\\
{\sf To appear in: Diffuse Waves in Complex Media, edited by J.P. Fouque,\\
NATO ASI Series (Kluwer, Dordrecht, 1999).}
\end{abstract}
\newpage

\section{Introduction}
\label{intro}

The name ``random laser'' made its appearance a few years ago \cite{Wie95a}, in
connection with experiments on amplifying random media \cite{Law94}. The
concept goes back to Letokhov's 1967 proposal to use a mirrorless laser as an
optical frequency standard \cite{Let67}. Laser action requires gain and
feedback. In any laser, gain results from stimulated emission of radiation by
atoms in a non-equilibrium state. The random laser differs from a conventional
laser in that the feedback is provided by multiple scattering from disorder
rather than by confinement from mirrors. Because of the randomness, there is no
geometry-dependent shift of the laser line with respect to the atomic
transition frequency (hence the potential as a frequency standard). Stellar
atmospheres may form a naturally occuring realization of a random laser
\cite{Lav75}.

Possible applications as ``paint-on lasers'' \cite{PS94} have sparked an
intensive experimental and theoretical investigation of the interplay of
multiple scattering and stimulated emission. The topic has been reviewed by
Wiersma and Lagendijk (see Ref.\ \cite{Wie97} and these Proceedings). A
particularly instructive experiment \cite{Wie95b} was the demonstration of the
narrowing of the coherent backscattering cone as a result of stimulated
emission below the laser threshold. This experiment can be explained within the
framework of {\em classical wave\/} optics. {\em Wave\/} optics, as opposed to
ray optics, because coherent backscattering is an interference effect. {\em
Classical\/} optics, as opposed to quantum optics, because stimulated emission
can be described by a classical wave equation. (What is needed is a dielectric
constant with a negative imaginary part.)

In a recent work \cite{Bee98} we went beyond classical optics by studying the
photo\-detection statistics of amplified spontaneous emission from a random
medium. Spontaneous emission, as opposed to stimulated emission, is a quantum
optical phenomenon that can not be described by a classical wave equation. In
this contribution we review our theory, with several extensions (notably in
Secs.\ \ref{onemode}, \ref{waveguidegeometry}, \ref{randomlaser}, and
\ref{broadband}).

We start out in Sec.\ \ref{quantization} with a discussion of the quantization
of the electromagnetic field in absorbing or amplifying media. There exists a
variety of approaches to this problem  [9--17]. We will use the method of
input--output relations developed by Gruner and Welsch \cite{Gru96a,Gru96b},
and by Loudon and coworkers \cite{Jef93,Bar95,Mat95,Mat97}. The central formula
of this section is a fluctuation-dissipation relation, that relates the
commutator of the operators describing the quantum fluctuations in the
electromagnetic field to the deviation $\openeen -S\cdot S^{\dagger}$ from
unitarity of the scattering matrix $S$ of the system. The relation holds both
for absorbing and amplifying media. The absorbing medium is in thermal
equilibrium at temperature $T$, and expectation values can be computed in terms
of the Bose-Einstein function
\begin{equation}
f(\omega,T)=[\exp(\hbar\omega/k_{\rm B}T)-1]^{-1}.\label{BEfunction}
\end{equation}
The amplifying medium is not in thermal equilibrium, but the expectation values
can be obtained from those in the absorbing medium by evaluating the
Bose-Einstein function at a negative temperature \cite{Jef93,Mat97}.

In Sec.\ \ref{photodetection} we apply this general framework to a
photo\-detection measurement. Our central result is a relationship between the
probability distribution $P(n)$ to count $n$ photons in a long time $t$ (long
compared to the coherence time of the radiation) and the eigenvalues
$\sigma_{1},\sigma_{2},\ldots\sigma_{N}$ of the matrix product $S\cdot
S^{\dagger}$. We call these eigenvalues ``scattering strengths''. They are
between $0$ and $1$ for an absorbing medium and greater than $1$ for an
amplifying medium. The mean photo\-count $\bar{n}$ is proportional to the
spectral average $N^{-1}\sum_{n}(1-\sigma_{n})$ of the scattering strengths.
This spectral average is the absorptivity of the medium, being the fraction of
the radiation incident in $N$ modes that is absorbed. (An amplifying medium has
a negative absorptivity.) The relation between mean photo\-count and
absorptivity constitutes Kirchhoff's law of thermal radiation. We generalize
Kirchhoff's law to higher moments of the counting distribution by relating the
$p$-th factorial cumulant of $n$ to $N^{-1}\sum_{n}(1-\sigma_{n})^{p}$. While
the absorptivity ($p=1$) can be obtained from the radiative transfer equation,
the spectral averages with $p>1$ can not. Fortunately, random-matrix theory
provides a set of powerful tools to compute such spectral averages \cite{RMP}.

We continue in Sec.\ \ref{applications} with the application of our formula for
the photo\-detection distribution to specific random media. We focus on two
types of geometries: An open-ended waveguide and a cavity containing a small
opening. Randomness is introduced by disorder or (in the case of the cavity) by
an irregular shape of the boundaries. Radiation is emitted into $N$ propagating
modes, which we assume to be a large number. (In the case of the cavity, $N$ is
the number of transverse modes in the opening.) It is unusual, but essential,
that all the emitted radiation is incident on the photo\-detector. We show that
if only a single mode is detected, the counting distribution contains solely
information on the absorptivity, while all information on higher spectral
moments of the scattering strengths is lost. To characterize the fluctuations
in the photo\-count we compute the variance ${\rm
Var}\,n=\overline{n^{2}}-\bar{n}^{2}$. The variance can be directly measured
from the auto-correlator of the photo\-current $I(t)=\bar{I}+\delta I(t)$,
according to
\begin{equation}
\int_{-\infty}^{\infty}dt\,\overline{\delta I(0)\delta
I(t)}=\lim_{t\rightarrow\infty}\frac{1}{t}{\rm Var}\,n.\label{khinchin}
\end{equation}
The bar $\overline{\cdots}$ indicates an average over many measurements on the
same sample. The mean photo\-count $\bar{n}$ (and hence the mean current
$\bar{I}=\bar{n}/t$) contains information on the absorptivity. The new
information contained in the variance of $n$ (or the auto-correlator of $I$) is
the effective number of degrees of freedom $\nu_{\rm eff}$, defined by
\cite{Man95} ${\rm Var}\,n=\bar{n}(1+\bar{n}/\nu_{\rm eff})$. For black-body
radiation in a narrow frequency interval $\delta\omega$, one has $\nu_{\rm
eff}=Nt\delta\omega/2\pi\equiv\nu$. The counting distribution is then a
negative-binomial distribution with $\nu$ degrees of freedom,
\begin{equation}
P(n)\propto{n+\nu-1\choose n}\exp(-n\hbar\omega/k_{\rm B}T).\label{binomial}
\end{equation}
The quantity $\nu_{\rm eff}$ generalizes the notion of degrees of freedom to
radiation from systems that are not black bodies.

A black body has scattering matrix $S=0$. (Any incident radiation is fully
absorbed.) In other words, the scattering strengths $\sigma_{n}$ of a black
body are all equal to zero. A random medium, in contrast, has in general a
broad (typically bimodal) density of scattering strengths. We show that this
results in a substantial reduction of $\nu_{\rm eff}$ below $\nu$. In other
words, the noise in the photo\-count is anomalously large in a random medium.
The reduction in $\nu_{\rm eff}$ holds both for absorbing and amplifying media.
The only requirement is a broad distribution of scattering strengths. For the
random laser, we predict that the ratio $\nu_{\rm eff}/\nu$ vanishes on
approaching the laser threshold. No such reduction is expected in a
conventional laser. We discuss the origin of this difference in the concluding
Sec.\ \ref{conclusion}, together with a discussion of the relationship between
$\nu_{\rm eff}$ and the Thouless number of mesoscopic physics.

\section{Quantization of the electromagnetic field}
\label{quantization}

\subsection{Input--output relations}
\label{inputoutput}

\begin{figure}
\hspace*{\fill}
\psfig{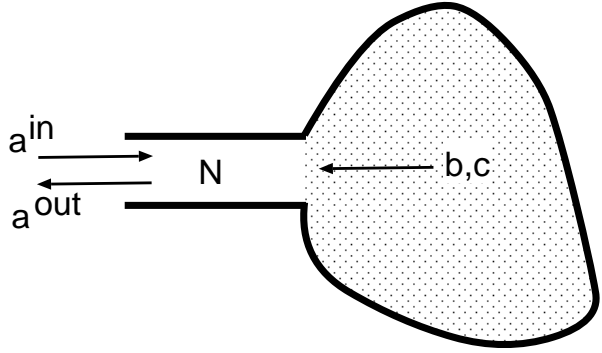}
\hspace*{\fill}
\caption[]{
Scattering geometry consisting of a random medium (dotted) coupled to free
space via an $N$-mode wave\-guide. The $N$-component vector of outgoing-mode
operators $a^{\rm out}$ is linearly related to the incoming-mode operators
$a^{\rm in}$ and the spontaneous-emission operators $b,c$.
\label{diagram2}
}
\end{figure}

We consider a dielectric medium coupled to free space via a wave\-guide with
$N(\omega)$ propagating modes (counting polarizations) at frequency $\omega$
(see Fig.\ \ref{diagram2}). The incoming and outgoing modes in the wave\-guide
are represented by two $N$-component vectors of annihilation operators $a^{\rm
in}(\omega)$, $a^{\rm out}(\omega)$. They satisfy the canonical commutation
relations
\begin{eqnarray}
&&[a_{n}^{{\rm in}\vphantom{\dagger}}(\omega),a_{m}^{{\rm
in}\dagger}(\omega')]= \delta_{nm}\delta(\omega-\omega'),\;\;
[a_{n}^{\rm in}(\omega),a_{m}^{\rm in}(\omega')]=0,\label{crain}\\
&&[a_{n}^{{\rm out}\vphantom{\dagger}}(\omega),a_{m}^{{\rm
out}\dagger}(\omega')]= \delta_{nm}\delta(\omega-\omega'),\;\;
[a_{n}^{\rm out}(\omega),a_{m}^{\rm out}(\omega')]=0.\label{craout}
\end{eqnarray}
The input--output relations take the form
\begin{equation}
a^{\rm out}=S\cdot a^{\rm in}+U\cdot b+V\cdot c^{\dagger}.\label{aoutSain}
\end{equation}
The two sets of operators $b,b^{\dagger}$ and $c,c^{\dagger}$ commute with each
other and with the set of input operators $a^{\rm in},a^{{\rm in}\dagger}$.
They satisfy the canonical commutation relations
\begin{eqnarray}
&&[b_{n}^{\vphantom{\dagger}}(\omega),b_{m}^{\dagger}(\omega')]=
\delta_{nm}\delta(\omega-\omega'),\;\;
[b_{n}(\omega),b_{m}(\omega')]=0,\label{crb}\\
&&[c_{n}^{\vphantom{\dagger}}(\omega),c_{m}^{\dagger}(\omega')]=
\delta_{nm}\delta(\omega-\omega'),\;\;
[c_{n}(\omega),c_{m}(\omega')]=0,\label{crc}
\end{eqnarray}
provided the $N\times N$ matrices $U(\omega)$ and $V(\omega)$ are related to
the scattering matrix $S(\omega)$ by
\begin{equation}
U\cdot U^{\dagger}-V\cdot V^{\dagger}=\openeen-S\cdot
S^{\dagger}\label{UVrelation}
\end{equation}
($\openeen$ denoting the $N\times N$ unit matrix). Equation (\ref{UVrelation})
can be understood as a fluctuation-dissipation relation: The left-hand side
accounts for quantum fluctuations in the electromagnetic field due to
spontaneous emission or absorption of photons, the right-hand side accounts for
dissipation due to absorption (or stimulated emission in the case of an
amplifying medium). Equation (\ref{UVrelation}) also represents a link between
classical optics (the scattering matrix $S$) and quantum optics (the quantum
fluctuation matrices $U,V$).

The matrix $\openeen-S\cdot S^{\dagger}$ is positive definite in an absorbing
medium, so we can put $V=0$ and write
\begin{equation}
a^{\rm out}=S\cdot a^{\rm in}+U\cdot b,\;\;U\cdot U^{\dagger}=\openeen-S\cdot
S^{\dagger}.\label{Vis0}
\end{equation}
(These are the input--output relations of Ref.\ \cite{Gru96b}.) Conversely, in
an amplifying medium $\openeen-S\cdot S^{\dagger}$ is negative definite, so we
can put $U=0$ and write
\begin{equation}
a^{\rm out}=S\cdot a^{\rm in}+V\cdot c^{\dagger},\;\;V\cdot V^{\dagger}=S\cdot
S^{\dagger}-\openeen.\label{Uis0}
\end{equation}
(The operator $c$ represents the inverted oscillator of Ref.\ \cite{Jef93}.)
Both matrices $U,V$ and operators $b,c$ are needed if $\openeen-S\cdot
S^{\dagger}$ is neither positive nor negative definite, which might happen if
the medium contains both absorbing and amplifying regions. In what follows we
will not consider that case any further.

\subsection{Expectation values}
\label{expectation}

We assume that the absorbing medium is in thermal equilibrium at temperature
$T$. Thermal emission is described by the operator $b$ with expectation value
\begin{equation}
\langle b_{n}^{\dagger}(\omega)b_{m}^{\vphantom{\dagger}}(\omega')\rangle=
\delta_{nm}\delta(\omega-\omega')f(\omega,T),\label{meanb}
\end{equation}
where $f$ is the Bose-Einstein function (\ref{BEfunction}).

The inverted oscillator $c$ accounts for spontaneous emission in an amplifying
medium. We consider the regime of linear amplification, below the laser
threshold. Formally, this regime can be described by a thermal distribution at
an effective {\em negative\/} temperature $-T$ \cite{Jef93,Mat97}. For a
two-level atomic system, with level spacing $\hbar\omega_{0}$ and an average
occupation $N_{\rm upper}>N_{\rm lower}$ of the two levels, the effective
temperature is given by $N_{\rm upper}/N_{\rm
lower}=\exp(\hbar\omega_{0}/k_{\rm B}T)$. The zero-temperature limit
corresponds to a complete population inversion. The expectation value is given
by
\begin{equation}
\langle c_{n}^{\vphantom{\dagger}}(\omega)c_{m}^{\dagger}(\omega')\rangle=
-\delta_{nm}\delta(\omega-\omega')f(\omega,-T),\label{meanc1}
\end{equation}
or equivalently by
\begin{equation}
\langle c_{n}^{\dagger}(\omega)c_{m}^{\vphantom{\dagger}}(\omega')\rangle=
\delta_{nm}\delta(\omega-\omega')f(\omega,T).\label{meanc2}
\end{equation}
(We have used that $f(\omega,T)+f(\omega,-T)=-1$.)

Higher order expectation values are obtained by pairwise averaging, as one
would do for Gaussian variables, after having brought the operators into normal
order (all creation operators to the left of the annihilation operators). This
procedure is an example of the ``optical equivalence theorem''
\cite{Man95,Lou83}. To do the Gaussian averages it is convenient to discretize
the frequency as $\omega_{p}=p\Delta$, $p=1,2,\ldots$, and send $\Delta$ to
zero at the end. The expectation value of an arbitrary functional ${\cal F}$ of
the operators $b,b^{\dagger}$ (or $c,c^{\dagger}$) can then be written as a
multiple integral over an array of complex numbers $z_{np}$ [$1\leq n\leq
N(\omega_{p})$],
\begin{equation}
\langle\,:{\cal
F}[\{b_{n}^{\vphantom{\dagger}}(\omega_{p})\},
\{b_{n}^{\dagger}(\omega_{p})\}]:\,\rangle =Z^{-1}
\int dz\,{\rm e}^{-\Phi}{\cal F}[\{z_{np}^{\vphantom{\ast}}\},
\{z_{np}^{\ast}\}], \label{calFaverage}
\end{equation}
with the definitions
\begin{eqnarray}
&&\Phi=\sum_{n,p}\frac{|z_{np}|^{2}\Delta}{f(\omega_{p},T)},\label{Phidef}\\
&&Z=\int dz\,{\rm e}^{-\Phi}=\prod_{n,p}\frac{\pi
f(\omega_{p},T)}{\Delta}.\label{Zdef}
\end{eqnarray}
The colons in Eq.\ (\ref{calFaverage}) indicate normal ordering, and $\int dz$
indicates the integration over the real and imaginary parts of all the
$z_{np}$'s.

\section{Photodetection statistics}
\label{photodetection}

\subsection{General formulas}
\label{general}

We consider the case that the incoming radiation is in the vacuum state, while
the outgoing radiation is collected by a photo\-detector. We assume a mode and
frequency independent detection efficiency of $\alpha$ photo\-electrons per
photon. The probability that $n$ photons are counted in a time $t$ is given by
the Glauber-Kelley-Kleiner formula \cite{Gla63,Kel64}
\begin{eqnarray}
&&P(n)=\frac{1}{n!}\langle\,:I^{n}\,{\rm
e}^{-I}:\,\rangle,\label{PIarelation}\\
&&I=\alpha\int_{0}^{t}dt'\,a^{{\rm out}\dagger}(t')\cdot a^{\rm
out}(t'),\label{Udef}
\end{eqnarray}
where we have defined the Fourier transform
\begin{equation}
a^{\rm out}(t)=(2\pi)^{-1/2}\int_{0}^{\infty}d\omega\,{\rm e}^{-{\rm i}\omega
t}a^{\rm out}(\omega).\label{aouttdef}
\end{equation}

The factorial cumulants $\kappa_{p}$ of $P(n)$ are the cumulants of the
factorial moments $\overline{n(n-1)\cdots(n-p+1)}$. For example,
$\kappa_{1}=\bar{n}$ and $\kappa_{2}=\overline{n(n-1)}-\bar{n}^{2}={\rm
Var}\,n-\bar{n}$. The factorial cumulants have the generating function
\begin{equation}
F(\xi)=\sum_{p=1}^{\infty}\frac{\kappa_{p}\xi^{p}}{p!}=
\ln\left(\sum_{n=0}^{\infty}(1+\xi)^{n}P(n)\right).\label{Fxidef1}
\end{equation}
Once $F(\xi)$ is known, the distribution $P(n)$ can be recovered from
\begin{equation}
P(n)=\frac{1}{2\pi{\rm i}}\oint_{|z|=1}dz\,z^{-n-1}{\rm
e}^{F(z-1)}=\lim_{\xi\rightarrow -1}\frac{1}{n!}\frac{d^{n}}{d\xi^{n}}{\rm
e}^{F(\xi)}.\label{PfromF}
\end{equation}
{}From Eq.\ (\ref{PIarelation}) one finds the expression
\begin{equation}
{\rm e}^{F(\xi)}=\langle\,:{\rm e}^{\xi I}:\,\rangle.\label{Fxidef2}
\end{equation}

To evaluate Eq.\ (\ref{Fxidef2}) for the case of an absorbing medium, we
combine Eq.\ (\ref{Vis0}) with Eqs.\ (\ref{Udef}) and (\ref{aouttdef}), and
then compute the expectation value with the help of Eq.\ (\ref{calFaverage}),
\begin{eqnarray}
&&{\rm e}^{F(\xi)}=Z^{-1}\int
dz\,\exp\left(-\Delta\sum_{n\vphantom{n'},p\vphantom{p'}}
\sum_{n',p'}z^{\ast}_{np\vphantom{n'}} M^{\vphantom{\ast}}_{np,n'p'}
z^{\vphantom{\ast}}_{n'p'}\right),
\label{Fxigeneral}\\
&&M_{np,n'p'}=\frac{\delta_{nn'}\delta_{pp'}}{f(\omega_{p},T)}
-\frac{\xi\alpha\Delta}{2\pi}\int_{0}^{t}dt'\,{\rm e}^{{\rm
i}(\omega_{p}-\omega_{p'})t'}\sum_{m}U^{\dagger}_{nm}(\omega_{p})
U^{\vphantom{\dagger}}_{mn'}(\omega_{p'}).\nonumber\\
&&\mbox{}\label{Mdef}
\end{eqnarray}
Evaluation of the Gaussian integrals results in the compact expression
\begin{equation}
F(\xi)={\rm constant}-\ln\|M\|,\label{Fxigeneralresult}
\end{equation}
where $\|\cdots\|$ indicates the determinant. (The $\xi$-independent constant
can be found from the normalization requirement that $F(0)=0$.) The matrix $U$
is related to the scattering matrix $S$ by $U\cdot U^{\dagger}=\openeen -S\cdot
S^{\dagger}$ [Eq.\ (\ref{Vis0})]. This relation determines $U$ up to a
transformation $U\rightarrow U\cdot A$, with $A(\omega)$ an arbitrary unitary
matrix. Since the determinant $\|M\|$ is invariant under this transformation,
we can say that knowledge of the scattering matrix suffices to determine the
counting distribution.

The result for an amplifying medium is also given by Eqs.\ (\ref{Mdef}) and
(\ref{Fxigeneralresult}), with the replacement of $U$ by $V$ and
$f(\omega_{p},T)$ by $-f(\omega_{p},-T)$ [in accordance with Eqs.\ (\ref{Uis0})
and (\ref{meanc1})].

The determinant $\|M\|$ can be simplified in the limit of large and small
counting times $t$. These two regimes will be discussed separately in the next
two subsections. A simple expression valid for all $t$ exists for the mean
photo\-count,
\begin{equation}
\bar{n}=t\int_{0}^{\infty}\!\frac{d\omega}{2\pi}\,\alpha f\,{\rm
Tr}\,(\openeen-S\cdot S^{\dagger}).\label{Kirchhoff}
\end{equation}
The quantity $N^{-1}{\rm Tr}\,(\openeen-S\cdot S^{\dagger})$ is the
absorptivity, defined as the fraction of the incident power that is absorbed at
a certain frequency, averaged over all incoming modes. The relation
(\ref{Kirchhoff}) between thermal emission and absorption is {\em Kirchhoff's
law}. It holds also for an amplifying medium, upon replacement of $f(\omega,T)$
by $f(\omega,-T)$.\footnote{
Eq.\ (\protect\ref{Fxigeneralresult}) for an amplifying system is obtained by
the replacement of $U$ by $V$ and $f(\omega,T)$ by {\em minus\/}
$f(\omega,-T)$. Since $V\cdot V^{\dagger}$ equals {\em minus\/}
$\openeen-S\cdot S^{\dagger}$ [Eq.\ (\protect\ref{Uis0})], the two minus signs
cancel and the net result for Eq.\ (\protect\ref{Kirchhoff}) is that we should
replace $f(\omega,T)$ by {\em plus\/} $f(\omega,-T)$.}

\subsection{Long-time regime}
\label{longtimeregime}

The long-time regime is reached when $\omega_{\rm c}t\gg 1$, with $\omega_{\rm
c}$ the frequency interval within which $S\cdot S^{\dagger}$ does not vary
appreciably. In this regime we may choose the discretization
$\omega_{p}=p\Delta$, $\Delta=2\pi/t$, satisfying
\begin{equation}
\int_{0}^{t}dt'\,{\rm e}^{{\rm
i}(\omega_{p}-\omega_{p'})t'}=t\delta_{pp'}.\label{discretization}
\end{equation}
The matrix (\ref{Mdef}) then becomes diagonal in the indices $p,p'$,
\begin{equation}
M_{np,n'p'}=\frac{\delta_{nn'}\delta_{pp'}}{f(\omega_{p},T)}
-\xi\alpha\delta_{pp'}\biggl(U^{\dagger}(\omega_{p})\cdot
U^{\vphantom{\dagger}}(\omega_{p})\biggr)_{nn'},\label{Mlong}
\end{equation}
so that the generating function (\ref{Fxigeneralresult}) takes the form
\begin{eqnarray}
F(\xi)&=&-t\int_{0}^{\infty}\!\frac{d\omega}{2\pi}\,
\ln\bigl\|\openeen-(\openeen-S\cdot S^{\dagger})\xi\alpha f\bigr\|\nonumber\\
&=&-t\int_{0}^{\infty}\!\frac{d\omega}{2\pi}\sum_{n=1}^{N(\omega)}
\ln\bigl[1-[1-\sigma_{n}(\omega)]\xi\alpha f(\omega,T)\bigr].\label{Fxilong}
\end{eqnarray}
We have introduced the {\em scattering strengths\/}
$\sigma_{1},\sigma_{2},\ldots\sigma_{N}$, being the eigenvalues of the
scattering-matrix product $S\cdot S^{\dagger}$. The result (\ref{Fxilong})
holds also for an amplifying system, if we replace $f(\omega,T)$ by
$f(\omega,-T)$.

Expansion of the logarithm in powers of $\xi$ yields the factorial cumulants
[cf.\ Eq.\ (\ref{Fxidef1})]
\begin{equation}
\kappa_{p}=(p-1)!\,t\int_{0}^{\infty}\!\frac{d\omega}{2\pi}(\alpha
f)^{p}\sum_{n=1}^{N}(1-\sigma_{n})^{p}.\label{kapparesult}
\end{equation}
The $p$-th factorial cumulant of $P(n)$ is proportional to the $p$-th spectral
moment of the scattering strengths. It is a special property of the long-time
regime that the counting distribution is determined entirely by the eigenvalues
of $S\cdot S^{\dagger}$, independently of the eigenfunctions. Eq.\
(\ref{kapparesult}) can be interpreted as a generalization of Kirchhoff's law
(\ref{Kirchhoff}) to higher moments of the counting distribution.

\subsection{Short-time regime}
\label{shorttimeregime}

The short-time regime is reached when $\Omega_{\rm c}t\ll 1$, with $\Omega_{\rm
c}$ the frequency range over which $S\cdot S^{\dagger}$ differs appreciably
from the unit matrix. (The reciprocal of $\Omega_{\rm c}$ is the coherence time
of the thermal emissions.) In this regime we may replace $\exp[{\rm
i}(\omega_{p}-\omega_{p'})t']$ in Eq.\ (\ref{Mdef}) by 1, so that $M$
simplifies to
\begin{equation}
M_{np,n'p'}=\frac{\delta_{nn'}\delta_{pp'}}{f(\omega_{p},T)}
-\frac{t\xi\alpha\Delta}{2\pi}\sum_{m}U^{\dagger}_{nm}(\omega_{p})
U^{\vphantom{\dagger}}_{mn'}(\omega_{p'}).\label{Mshort1}
\end{equation}
If we suppress the mode indices $n,n'$, Eq.\ (\ref{Mshort1}) can be written as
\begin{equation}
M_{p,p'}=\frac{\delta_{pp'}}{f(\omega_{p},T)}\openeen
-\frac{t\xi\alpha\Delta}{2\pi}U^{\dagger}(\omega_{p})\cdot
U^{\vphantom{\dagger}}(\omega_{p'}).\label{Mshort2}
\end{equation}
The determinant $\|M_{pp'}\|$ can be evaluated with the help of the
formula\footnote{
To verify Eq.\ (\protect\ref{detformula}), take the logarithm of each side and
use $\ln\|M\|={\rm Tr}\,\ln M$. Then expand each logarithm in powers of $A$ and
equate term by term. I am indebted to J.M.J. van Leeuwen for helping me with
this determinant.
}
\begin{equation}
\|\delta_{pp'}\openeen+A_{p\vphantom{p'}}\cdot
B_{p'}\|=\bigl\|\openeen+\sum_{q}B_{q}\cdot A_{q}\bigr\| \label{detformula}
\end{equation}
(with $\{A_{p}\}$, $\{B_{p}\}$ two arbitrary sets of matrices). The resulting
generating function is\footnote{
We adopt the convention that the matrix $\openeen-S\cdot S^{\dagger}$ is
embedded in an infinite-dimensional matrix by adding zeroes. The matrix
$\openeen$ outside the integral over $\omega$ in Eq.\ (\protect\ref{Fxishort})
is then interpreted as an infinite-dimensional unit matrix.
}
\begin{equation}
F(\xi)=-\ln\bigl\|\openeen-t\int_{0}^{\infty}\!
\frac{d\omega}{2\pi}\,(\openeen-S\cdot S^{\dagger})\xi\alpha f\bigr\|.
\label{Fxishort}
\end{equation}
Again, to apply Eq.\ (\ref{Fxishort}) to an amplifying system we need to
replace $f(\omega,T)$ by $f(\omega,-T)$.

The short-time limit (\ref{Fxishort}) is more complicated than the long-time
limit (\ref{Fxilong}), in the sense that the former depends on both the
eigenvalues and eigenvectors of $S\cdot S^{\dagger}$. There is therefore no
direct relation between factorial cumulants of $P(n)$ and spectral moments of
scattering strengths in the short-time regime.

\subsection{Single-mode detection}
\label{onemode}

\begin{figure}
\hspace*{\fill}
\psfig{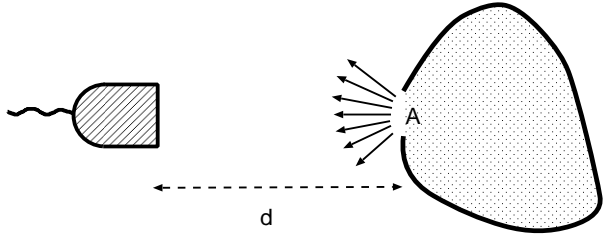}
\hspace*{\fill}
\caption[]{
A cavity radiates through a hole with an area ${\cal A}$. The number $N$ of
radiating modes at wavelength $\lambda$ is $2\pi{\cal A}/\lambda^{2}$ (counting
polarizations). All $N$ modes are detected if the photo\-cathode covers the
hole. Upon increasing the separation $d$ between hole and photo\-cathode, fewer
and fewer modes are detected. Finally, single-mode detection is reached when
the area of the photo\-cathode becomes less than the coherence area $\simeq
d^{2}/N$ of the radiation.
\label{singlemode}
}
\end{figure}

We have assumed that each of the $N$ radiating modes is detected with equal
efficiency $\alpha$. At the opposite extreme, we could assume that only a
single mode is detected. This would apply if the photo\-cathode had an area
smaller than the coherence area of the emitted radiation (see Fig.\
\ref{singlemode}). Single-mode detection is less informative than multi-mode
detection, for the following reason.

Suppose that only a single mode is detected. The counting distribution $P(n)$
is still given by Eq.\ (\ref{PIarelation}), but now $I$ contains only a single
element (say, number $1$) of the vector of operators $a^{\rm out}$,
\begin{equation}
I=\alpha\int_{0}^{t}dt'\,a^{{\rm out}\dagger}_{1}(t')a^{\rm
out}_{1}(t').\label{U1def}
\end{equation}
This amounts to the replacement of the matrix $U$ in Eq.\ (\ref{Mdef}) by the
projection ${\cal P}\cdot U$, with ${\cal P}=\delta_{nm}\delta_{n1}$. Instead
of Eq.\ (\ref{Kirchhoff}) we have the mean photo\-count
\begin{equation}
\bar{n}=\int_{0}^{\infty}d\omega\,\frac{d\bar{n}}{d\omega}
;\;\;\;\frac{d\bar{n}}{d\omega}= \frac{t\alpha f}{2\pi}[1-(S\cdot S^{\dagger})_{11}].
\label{Kirchhoff1}
\end{equation}
The generating function now takes the form
\begin{equation}
F(\xi)=-t\int_{0}^{\infty}\!\frac{d\omega}{2\pi}\ln\left(1-\frac{2\pi\xi}{t}
\frac{d\bar{n}}{d\omega}\right)\label{Fxi1long}
\end{equation}
in the long-time regime, and
\begin{equation}
F(\xi)=-\ln(1-\xi\bar{n})\label{Fxi1short}
\end{equation}
in the short-time regime. We see that the entire counting distribution is
determined by the mean photo\-count, hence by the absorptivity of the single
detected mode. This strong version of Kirchhoff's law is due to Bekenstein and
Schiffer \cite{Bek94}. It holds only for the case of single-mode detection.
Multi-mode detection is determined not just by $\bar{n}$, being the first
spectral moment of the scattering strengths, but also by higher moments.

\subsection{Waveguide geometry}
\label{waveguidegeometry}

\begin{figure}
\hspace*{\fill}
\psfig{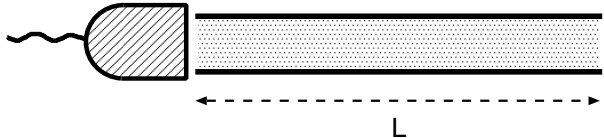}
\hspace*{\fill}
\caption[]{
Disordered wave\-guide (length $L$) connected at one end to a photo\-detector.
\label{figwaveguide}
}
\end{figure}

Figs.\ \ref{diagram2} and \ref{singlemode} show a cavity geometry.
Alternatively, one can consider the waveguide geometry of Fig.\
\ref{figwaveguide}. The waveguide has cross-section ${\cal A}$, corresponding
to $N=2\pi{\cal A}/\lambda^{2}$ modes at wavelength $\lambda$. The $2N\times
2N$ scattering matrix $S$ consists of four $N\times N$ blocks,
\begin{equation}
S=\left(\begin{array}{cc}
r&t\\t'&r'
\end{array}\right),\label{Srt}
\end{equation}
namely two reflection matrices $r,r'$ (reflection from the left and from the
right) and two transmission matrices $t,t'$ (transmission from right to left
and from left to right). Reciprocity relates $t$ and $t'$ (they are each others
transpose).

A photo\-detector detects the radiation emitted at one end of the waveguide,
while the radiation emitted at the other end remains undetected. If the
radiation from {\em both\/} ends would be detected (by two photo\-detectors),
then the eigenvalues of $S\cdot S^{\dagger}$ would determine the counting
distribution in the long-time limit, as in the cavity geometry. But for
detection at the left end only, one needs instead the eigenvalues of the matrix
$r\cdot r^{\dagger}+t\cdot t^{\dagger}$ (or $r'\cdot r'^{\dagger}+t'\cdot
t'^{\dagger}$ for detection at the right end). More precisely, the general
expression for the characteristic function is given by Eqs.\ (\ref{Fxigeneral})
and (\ref{Mdef}) upon replacement of the $2N\times 2N$ matrix $U$ by the
projection ${\cal P}\cdot U$, with ${\cal P}_{nm}=1$ if $1\leq n=m\leq N$ and
${\cal P}_{nm}=0$ otherwise. In the long-time regime one obtains
\begin{equation}
F(\xi)=-t\int_{0}^{\infty}\!\frac{d\omega}{2\pi}
\,\ln\bigl\|\openeen-(\openeen-r\cdot r^{\dagger}-t\cdot t^{\dagger})
\xi\alpha f\bigr\|,\label{Fxilongwaveguide}
\end{equation}
and in the short-time regime
\begin{equation}
F(\xi)=-\ln\bigl\|\openeen-t\int_{0}^{\infty}\!
\frac{d\omega}{2\pi}\,(\openeen-r\cdot r^{\dagger}-t\cdot t^{\dagger})
\xi\alpha f\bigr\|.\label{Fxishortwaveguide}
\end{equation}

The eigenvalues of $r\cdot r^{\dagger}+t\cdot t^{\dagger}$ differ from the sum
$R_{n}+T_{n}$ of the reflection and transmission eigenvalues (eigenvalues of
$r\cdot r^{\dagger}$ and $t\cdot t^{\dagger}$, respectively), because the two
matrices $r\cdot r^{\dagger}$ and $t\cdot t^{\dagger}$ do not commute. The
absorptivity
\[
N^{-1}{\rm Tr}\,(\openeen-r\cdot r^{\dagger}-t\cdot
t^{\dagger})=N^{-1}\sum_{n=1}^{N}(1-R_{n}-T_{n})
\]
does depend only on the reflection and transmission eigenvalues. It determines
the mean photo\-count
\begin{equation}
\bar{n}=t\int_{0}^{\infty}\!\frac{d\omega}{2\pi}\,\alpha f\,{\rm
Tr}\,(\openeen-r\cdot r^{\dagger}-t\cdot
t^{\dagger}),\label{Kirchhoffwaveguide}
\end{equation}
in accordance with Kirchhoff's law. Higher moments of the counting distribution
can not be obtained from the reflection and transmission eigenvalues, but
require knowledge of the eigenvalues of $r\cdot r^{\dagger}+t\cdot
t^{\dagger}$. A substantial simplification occurs, in the case of absorption,
if the waveguide is sufficiently long that there is no transmission through it.
Then $t\cdot t^{\dagger}$ can be neglected and the counting distribution
depends entirely on the reflection eigenvalues.

\section{Applications}
\label{applications}

\subsection{Black-body radiation}
\label{blackbody}

Let us first check that we recover the familiar results for black-body
radiation \cite{Man95,Lou83}. The simplest ``step-function model'' of a black
body has
\begin{equation}
S(\omega)=\left\{\begin{array}{cc}
0&{\rm for}\;\;|\omega-\omega_{0}|<\frac{1}{2}\Omega_{\rm c},\\
\openeen&{\rm for}\;\;|\omega-\omega_{0}|>\frac{1}{2}\Omega_{\rm c}.
\end{array}\right.
\label{stepfunction}
\end{equation}
Incident radiation is fully absorbed within the frequency interval $\Omega_{\rm
c}$ around $\omega_{0}$ and fully reflected outside this interval. Typically,
$\Omega_{\rm c}\ll\omega_{0}$ so that we may neglect the frequency dependence
of $N(\omega)$ and $f(\omega,T)$, replacing these quantities by their values at
$\omega=\omega_{0}$.

The generating function (\ref{Fxilong}) in the long-time regime then becomes
\begin{equation}
F(\xi)=-\frac{Nt\Omega_{\rm c}}{2\pi}\ln(1-\xi\alpha f).\label{Fxilongbb}
\end{equation}
The inversion formula (\ref{PfromF}) yields the counting distribution
\begin{equation}
P(n)=\frac{\Gamma(n+\nu)}{n!\Gamma(\nu)}\frac{(\bar{n}/\nu)^{n}}
{(1+\bar{n}/\nu)^{n+\nu}}. \label{Pnlongbb}
\end{equation}
This is the negative-binomial distribution with $\nu=Nt\Omega_{\rm c}/2\pi$
degrees of freedom. [For integer $\nu$, the ratio of Gamma functions forms the
binomial coefficient ${n+\nu-1\choose n}$ that counts the number of partitions
of $n$ bosons among $\nu$ states, cf.\ Eq.\ (\ref{binomial}).] Note that
$\nu\gg 1$ in the long-time regime. The mean photo\-count is $\bar{n}=\nu\alpha
f$. In the limit $\bar{n}/\nu\rightarrow 0$, the negative-binomial distribution
tends to the Poisson distribution
\begin{eqnarray}
P(n)=\frac{1}{n!}\bar{n}^{n}{\rm e}^{-\bar{n}}\label{PPoisson}
\end{eqnarray}
of independent photo\-counts. The negative-binomial distribution describes
photo\-counts that occur in ``bunches''. Its variance
\begin{equation}
{\rm Var}\,n=\bar{n}(1+\bar{n}/\nu)\label{Varnbb}
\end{equation}
is larger than the Poisson value by a factor $1+\bar{n}/\nu$.

Similarly, the short-time limit (\ref{Fxishort}) becomes
\begin{equation}
F(\xi)=-N\ln\left(1-\frac{t\Omega_{\rm c}}{2\pi}\xi\alpha
f\right),\label{Fxishortbb}
\end{equation}
corresponding to a negative-binomial distribution with $N$ degrees of freedom.

In the step-function model (\ref{stepfunction}) $S$ changes abruptly from $0$
to $\openeen$ at $|\omega-\omega_{0}|=\frac{1}{2}\Omega_{\rm c}$. A more
realistic model would have a gradual transition. A Lorentzian frequency profile
is commonly used in the literature \cite{Meh70}, for the case of single-mode
detection. Substitution of $1-(S\cdot
S^{\dagger})_{11}=\bigl[1+4(\omega-\omega_{0})^{2}/\Omega_{\rm
c}^{2}\bigr]^{-1}$ into Eq.\ (\ref{Kirchhoff1}) and integration over $\omega$
in Eq.\ (\ref{Fxi1long}) (neglecting the frequency dependence of $f$) yields
the generating function in the long-time regime,
\begin{equation}
F(\xi)={\textstyle\frac{1}{2}}t\Omega_{\rm c}\left(1-\sqrt{1-\xi\alpha
f}\right).\label{FxilongbbL}
\end{equation}
The corresponding counting distribution is
\begin{equation}
P(n)=\frac{C}{n!}\left(\frac{\bar{n}}{\sqrt{1+\alpha f}}\right)^{n}
K_{n-1/2}\left({\textstyle\frac{1}{2}}t\Omega_{\rm c}\sqrt{1+\alpha
f}\right),\label{Pnbb}
\end{equation}
with $\bar{n}=\frac{1}{4}t\Omega_{\rm c}\alpha f$ and $K$ a Bessel function.
[The normalization constant is $C=\exp(\frac{1}{2}t\Omega_{\rm c})(t\Omega_{\rm
c}/\pi)^{1/2}(1+\alpha f)^{1/4}$.] This distribution was first obtained by
Glauber \cite{Gla63}. It is closely related to the socalled $K$-distribution in
the theory of scattering from turbulent media \cite{Jak78,Jak80}.
The counting distribution (\ref{Fxi1short}) in the short-time regime remains
negative-binomial.

In most realizations of black-body radiation the value of the Bose-Einstein
function $f(\omega_{0},T)$ is $\ll 1$. The difference between the two
distributions (\ref{Pnlongbb}) and (\ref{Pnbb}) is then quite small, both being
close to the Poisson distribution (\ref{PPoisson}).

\subsection{Reduction of degrees of freedom}
\label{reduction}

We now turn to applications of our general formulas to specific random media.
We concentrate on the long-time regime and assume a frequency-resolved
measurement, in which photons are only counted within a frequency interval
$\delta\omega$ around $\omega_{0}$. (For a black body, this would correspond to
the step-function model with $\Omega_{\rm c}$ replaced by $\delta\omega$.) We
take $\delta\omega$ smaller than any of the characteristic frequencies
$\omega_{\rm c}$, $\Omega_{\rm c}$, but necessarily greater than $1/t$. The
factorial cumulants are then given by
\begin{equation}
\kappa_{p}=(p-1)!\,\nu (\alpha
f)^{p}\frac{1}{N}\sum_{n=1}^{N}(1-\sigma_{n})^{p},\label{kapparesult2}
\end{equation}
where $\nu=Nt\delta\omega/2\pi$. For comparison with black-body radiation we
parameterize the variance in terms of the effective number $\nu_{\rm eff}$ of
degrees of freedom \cite{Man95},
\begin{equation}
{\rm Var}\,n=\bar{n}(1+\bar{n}/\nu_{\rm eff}),\label{nueffdef}
\end{equation}
with $\nu_{\rm eff}=\nu$ for a black body [cf.\ Eq.\ (\ref{Varnbb})]. Eq.\
(\ref{kapparesult2}) implies
\begin{equation}
\frac{\nu_{\rm
eff}}{\nu}=\frac{\bigl[\sum_{n}(1-\sigma_{n})\bigr]^{2}}{N\sum_{n}
(1-\sigma_{n})^{2}}\leq 1.\label{nueffrho}
\end{equation}
We conclude that the super-Poissonian noise of a random medium corresponds to a
black body with a {\em reduced\/} number of degrees of freedom. The reduction
occurs only for multi-mode emission. (Eq.\ (\ref{nueffrho}) with $N=1$ gives
$\nu_{\rm eff}=\nu$.) In addition, it requires multi-mode detection to observe
the reduction, because single-mode detection contains no other information than
the absorptivity (cf.\ Sec.\ \ref{onemode}).

An ensemble of random media has a certain scattering-strength density
\begin{equation}
\rho(\sigma)=\left\langle\sum_{n=1}^{N}
\delta(\sigma-\sigma_{n})\right\rangle,\label{rhodef}
\end{equation}
where the brackets $\langle\cdots\rangle$ denote the ensemble average. In the
large-$N$ regime sample-to-sample fluctuations are small, so the ensemble
average is representative for a single system.\footnote{
This statement is strictly speaking not correct for amplifying systems. The
reason is that the ensemble average is dominated by a small fraction of members
of the ensemble that are above the laser threshold, and this fraction is
non-zero for any non-zero amplification rate. This is a non-perturbative
finite-$N$ effect that does not appear if the ensemble average is computed
using the large-$N$ perturbation theory employed here.
}
We may therefore replace $\sum_{n}$ by $\int d\sigma\,\rho(\sigma)$ in Eqs.\
(\ref{kapparesult2}) and (\ref{nueffrho}). In the applications that follow we
will restrict ourselves to the large-$N$ regime, so that we can ignore
sample-to-sample fluctuations. All that we need in this case is the function
$\rho(\sigma)$. Random-matrix theory \cite{RMP} provides a method to compute
this function for a variety of random media.

\subsection{Disordered wave\-guide}
\label{waveguide}

As a first example we consider the thermal radiation from a disordered
absorbing wave\-guide (Fig.\ \ref{figwaveguide}). The length of the wave\-guide
is $L$, the transport mean free path in the medium is $l$, the velocity of
light is $c$, and $\tau_{\rm s}=l/c$ is the scattering time. The absorption
time $\tau_{\rm a}$ is related to the imaginary part $\varepsilon''>0$ of the
(relative) dielectric constant by $1/\tau_{\rm a}=\omega_{0}\varepsilon''$. We
assume that $\tau_{\rm s}$ and $\tau_{\rm a}$ are both $\gg 1/\omega_{0}$, so
that scattering as well as absorption occur on length scales large compared to
the wavelength. We define the normalized absorption rate\footnote{
The coefficient $16/3$ in Eq.\ (\protect\ref{gammadef}) is chosen to facilitate
the comparison between wave\-guide and cavity in the next subsection, and
refers to three-dimensional scattering. In the case of two-dimensional
scattering the coefficient is $\pi^{2}/2$. The present definition of $\gamma$
differs from that used in Ref.\ \protect\cite{Bee96} by a factor of two.
}
\begin{equation}
\gamma=\frac{16}{3}\frac{\tau_{\rm s}}{\tau_{\rm a}}.\label{gammadef}
\end{equation}
We call the system weakly absorbing if $\gamma\ll 1$, meaning that the
absorption is weak on the scale of the mean free path. If $\gamma\gg 1$ we call
the system strongly absorbing, $\gamma\rightarrow\infty$ being the black-body
limit. For simplicity we restrict ourselves to the case of an infinitely long
wave\-guide (more precisely, $L\gg l/\sqrt{\gamma}$), so that transmission
through it can be neglected.\footnote{
Results for an absorbing wave\-guide of finite length follow from Eqs.\
(\protect\ref{barnslab2})--(\protect\ref{nueffslab2}) upon changing the sign of
$\gamma$.}

\begin{figure}
\hspace*{\fill}
\psfig{figure=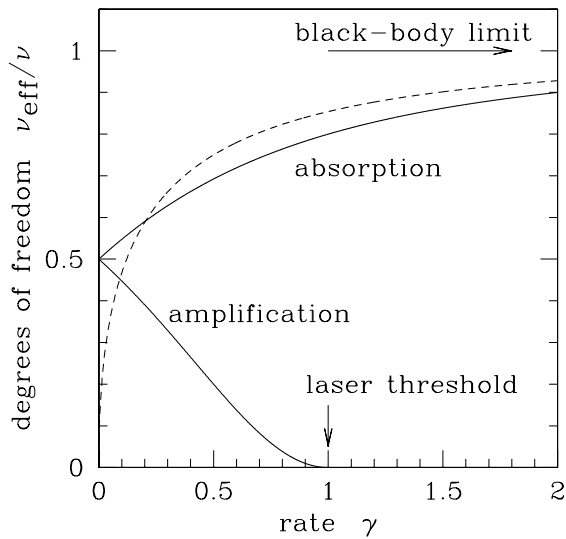,width=8cm}
\hspace*{\fill}
\caption[]{
Effective number of degrees of freedom as a function of normalized absorption
or amplification rate. The dashed curve is Eq.\ (\protect\ref{nueffslab}) for
an absorbing, infinitely long disordered waveguide, the solid curves are Eqs.\
(\protect\ref{nueffcavity}) and (\protect\ref{nueffcavity2}) for the chaotic
cavity. For the cavity both the cases of absorption and amplification are
shown. (See Fig. \protect\ref{figslab} for the amplifying wave\-guide.) The
black-body limit for absorbing systems and the laser threshold for amplifying
systems are indicated by arrows.
\label{nueff}
}
\end{figure}

In the absence of transmission the scattering matrix $S$ coincides with the
reflection matrix $r$, and the scattering strengths $\sigma_{n}$ coincide with
the reflection eigenvalues $R_{n}$ (eigenvalues of $r\cdot r^{\dagger}$). The
density $\rho(\sigma)$ for this system is known for any value of $N$
\cite{Bee96,Bru96}. The general expression is a series of Laguerre polynomials,
which in the large-$N$ regime of present interest simplifies to
\begin{equation}
\rho(\sigma)=\frac{N\sqrt{\gamma}}{\pi}
\frac{(\sigma^{-1}-1-{\textstyle\frac{1}{4}}\gamma)^{1/2}} 
{(1-\sigma)^{2}},\;\; 0<\sigma<\frac{1}{1+{\textstyle\frac{1}{4}}\gamma}.
\label{rhoslab}
\end{equation}
(The large-$N$ regime requires $N\gg 1$, but in weakly absorbing systems the
condition is stronger: $N\gg 1/\sqrt{\gamma}$.) This leads to the effective
number of degrees of freedom
\begin{equation}
\frac{\nu_{\rm eff}}{\nu}=\frac{\left[\int
d\sigma\,\rho(\sigma)(1-\sigma)\right]^{2}} {N\int
d\sigma\,\rho(\sigma)(1-\sigma)^{2}}=
4[(1+4/\gamma)^{1/4}+(1+4/\gamma)^{-1/4}]^{-2}, \label{nueffslab}
\end{equation}
plotted in Fig.\ \ref{nueff}, with a mean photo\-count of
\begin{equation}
\bar{n}={\textstyle\frac{1}{2}}\nu\alpha
f\gamma\left(\sqrt{1+4/\gamma}-1\right).\label{barnslab}
\end{equation}
For strong absorption, $\gamma\gg 1$, we recover the black-body result
$\nu_{\rm eff}=\nu$, as expected. For weak absorption, $\gamma\ll 1$, we find
$\nu_{\rm eff}=2\nu\sqrt{\gamma}$.

The characteristic function in the long-time frequency-resolved regime follows
from
\begin{equation}
F(\xi)=-\frac{\nu}{N}\int d\sigma\,\rho(\sigma) \ln\bigl[1-(1-\sigma)\xi\alpha
f\bigr].\label{Fxilong2}
\end{equation}
Substitution of Eq.\ (\ref{rhoslab}) into Eq.\ (\ref{Fxilong2}) yields a
hypergeometric function, which in the limit $\gamma\ll 1$ of weak absorption
simplifies to
\begin{equation}
F(\xi)=\nu_{\rm eff}\left(1-\sqrt{1-\xi\alpha f}\right),\;\;\nu_{\rm
eff}=2\nu\sqrt{\gamma}. \label{Fxislab}
\end{equation}
The counting distribution corresponding to Eq.\ (\ref{Fxislab}),
\begin{equation}
P(n)\propto\frac{1}{n!}\left(\frac{\bar{n}}{\sqrt{1+\alpha f}}\right)^{n}
K_{n-1/2}\left(\nu_{\rm eff}\sqrt{1+\alpha f}\right),\label{Pnslab}
\end{equation}
is Glauber's distribution (\ref{Pnbb}) with an effective number of degrees of
freedom. Note that Eq.\ (\ref{Pnbb}) resulted from single-mode detection over a
broad frequency range, whereas Eq.\ (\ref{Pnslab}) results from multi-mode
detection over a narrow frequency range. It appears as a coincidence that the
two distributions have the same functional form (with different parameters).

\subsection{Chaotic cavity}
\label{chaotic}

Our second example is an optical cavity radiating through a small hole covered
by a photo\-detector (Fig.\ \ref{singlemode}). The area ${\cal A}$ of the hole
should be small compared to the surface area of the cavity. The cavity should
have an irregular shape, or it should contain random scatterers --- to ensure
chaotic scattering of the radiation inside the cavity. It should be large
enough that the spacing $\Delta\omega$ of the cavity modes near frequency
$\omega_{0}$ is $\ll\omega_{0}$. For this system we define the normalized
absorption rate as
\begin{equation}
\gamma=\frac{\tau_{\rm dwell}}{\tau_{\rm a}};\;\;\;\tau_{\rm dwell}\equiv
\frac{2\pi}{N\Delta\omega}.\label{gammadef2}
\end{equation}
The time $\tau_{\rm dwell}$ is the mean dwell time of a photon in the cavity
without absorption. The frequency $1/\tau_{\rm dwell}$ represents the
broadening of the cavity modes due to the coupling to the $N=2\pi{\cal
A}/\lambda^{2}$ modes propagating through the hole. The broadening is much
greater than the spacing $\Delta\omega$ for $N\gg 1$. The large-$N$ regime
requires in addition $N\gg 1/\gamma$. The scattering-strength density in the
large-$N$ regime can be calculated using the perturbation theory of Ref.\
\cite{Bro96}.

\begin{figure}
\hspace*{\fill}
\psfig{figure=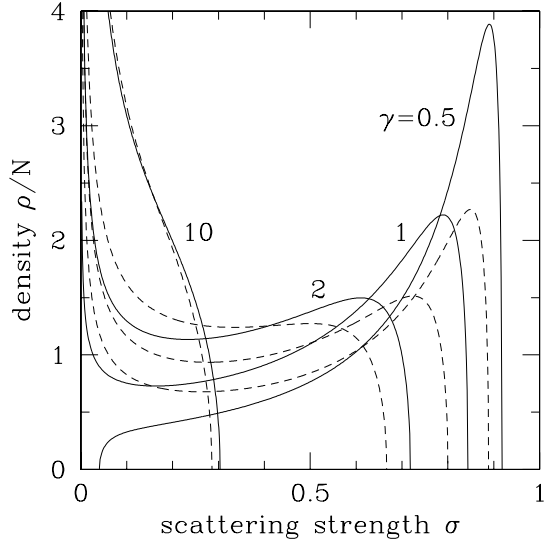,width=8cm}
\hspace*{\fill}
\caption[]{
Solid curves: Scattering-strength density of an absorbing chaotic cavity in the
large-$N$ regime, calculated from Eq.\ (\protect\ref{fullresult}) for four
values of the dimensionless absorption rate $\gamma$. The density
(\protect\ref{rhoslab}) for an absorbing, infinitely long disordered
wave\-guide is included for comparison (dashed). For $\gamma\gg 1$ the results
for cavity and wave\-guide coincide.
\label{rhoplot}
}
\end{figure}

The result is a rather complicated algebraic function, see the Appendix. It has
a simple form in the limit $\gamma\ll 1$ of weak absorption,
\begin{equation}
\rho(\sigma)=\frac{N}{2\pi}\frac{(\sigma-\sigma_{-})^{1/2}
(\sigma_{+}-\sigma)^{1/2}} {(1-\sigma)^{2}},\;\;\sigma_{-}<\sigma<\sigma_{+},
\label{rhocavity}
\end{equation}
with $\sigma_{\pm}=1-3\gamma\pm 2\gamma\sqrt{2}$. In the opposite limit
$\gamma\gg 1$ of strong absorption, $\rho(\sigma)$ is given by the same Eq.\
(\ref{rhoslab}) as for the infinitely long disordered wave\-guide. The
crossover from weak to strong absorption is shown in Fig.\ \ref{rhoplot}. The
value $\gamma=1$ is special in the sense that $\rho(\sigma)$ goes to zero or
infinity as $\sigma\rightarrow 0$, depending on whether $\gamma$ is smaller or
greater than 1.

We find the mean and variance of the photo\-count
\begin{eqnarray}
&&\bar{n}=\frac{\nu\alpha f\gamma}{1+\gamma},\label{barncavity}\\
&&{\rm Var}\,n=\bar{n}+\nu(\alpha
f)^{2}\gamma^{2}\frac{\gamma^{2}+2\gamma+2}{(1+\gamma)^{4}},\label{Varncavity}
\end{eqnarray}
corresponding to the effective number of degrees of freedom
\begin{equation}
\frac{\nu_{\rm eff}}{\nu}=\frac{(1+\gamma)^{2}}{\gamma^{2}+2\gamma+2}.
\label{nueffcavity}
\end{equation}
Again, $\nu_{\rm eff}=\nu$ for $\gamma\gg 1$. For $\gamma\ll 1$ we find
$\nu_{\rm eff}={\textstyle\frac{1}{2}}\nu$. This factor-of-two reduction of the
number of degrees of freedom is a ``universal'' result, independent of any
parameters of the system. The chaotic cavity is compared with the disordered
wave\-guide in Fig.\ \ref{nueff}. The ratio $\nu_{\rm eff}/\nu$ for the chaotic
cavity remains finite no matter how weak the absorption, while this ratio goes
to zero when $\gamma\rightarrow 0$ in the case of the infinitely long
disordered wave\-guide.\footnote{
For a wave\-guide of finite length $L$ ($\gg l$) one has instead $\nu_{\rm
eff}/\nu\rightarrow 5\,l/L$ in the limit $\gamma\rightarrow 0$ (cf.\ Sec.\
\protect\ref{randomlaser}).}

\subsection{Random laser}
\label{randomlaser}

The examples of the previous subsections concern thermal emission from
absorbing systems. As we discussed in Sec.\ \ref{longtimeregime}, our general
formulas can also be applied to amplified spontaneous emission, by evaluating
the Bose-Einstein function $f$ at a negative temperature \cite{Jef93,Mat97}.
Complete population inversion corresponds to $f=-1$. The amplification rate
$1/\tau_{\rm a}=\omega_{0}|\varepsilon''|$ should be so small that we are well
below the laser threshold, in order to stay in the regime of linear
amplification. The laser threshold occurs when the normalized amplification
rate $\gamma$ reaches a critical value $\gamma_{\rm c}$. (Sample-to-sample
fluctuations in the laser threshold \cite{Zyu95} are small in the large-$N$
regime.) For the cavity $\gamma_{\rm c}=1$. For the disordered wave\-guide one
has
\begin{equation}
\gamma_{\rm c}=\left(\frac{4\pi l}{3L}\right)^{2}\;\;{\rm if}\;\;L\gg
l.\label{gammacdef}
\end{equation}
Since $\gamma_{\rm c}\rightarrow 0$ in the limit $L\rightarrow\infty$, the
infinitely long wave\-guide is above the laser threshold no matter how weak the
amplification.

\begin{figure}
\hspace*{\fill}
\psfig{figure=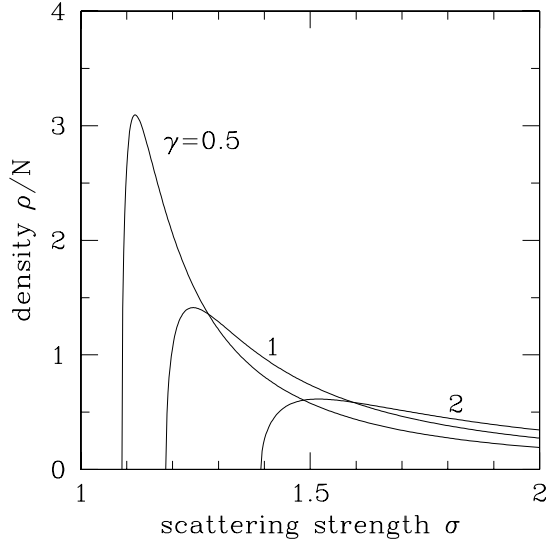,width=8cm}
\hspace*{\fill}
\caption[]{
Scattering-strength density of an amplifying chaotic cavity, calculated from
the results in Fig.\ \protect\ref{rhoplot} by means of the duality relation
(\protect\ref{rhopmrelation}). First and higher moments diverge for $\gamma\geq
1$.
\label{rhoplot2}
}
\end{figure}

A duality relation \cite{Paa96} between absorbing and amplifying systems
greatly simplifies the calculation of the scattering strengths. Dual systems
differ only in the sign of the imaginary part $\varepsilon''$ of the dielectric
constant (positive for the absorbing system, negative for the amplifying
system). Therefore, dual systems have the same value of $\tau_{\rm a}$ and
$\gamma$. The scattering matrices of dual systems are related by
$S_{-}^{\dagger}=S_{+}^{-1}$, hence $S_{-}^{\vphantom{\dagger}}\cdot
S_{-}^{\dagger}=(S_{+}^{\vphantom{\dagger}}\cdot S_{+}^{\dagger})^{-1}$. (The
subscript $+$ denotes the absorbing system, the subscript $-$ the dual
amplifying system.) We conclude that the scattering strengths
$\sigma_{1},\sigma_{2},\ldots\sigma_{N}$ of an amplifying system are the
reciprocal of those of the dual absorbing system. The densities
$\rho_{\pm}(\sigma)$ are related by
\begin{equation}
\sigma^{2}\rho_{-}(\sigma)=\rho_{+}(1/\sigma).\label{rhopmrelation}
\end{equation}
In Fig.\ \ref{rhoplot2} we show the result of the application of the
transformation (\ref{rhopmrelation}) to the densities of Fig.\ \ref{rhoplot}
for the case of a cavity. The critical value $\gamma_{\rm c}=1$ is such that
the first and higher moments are finite for $\gamma<\gamma_{\rm c}$ and
infinite for $\gamma\geq\gamma_{\rm c}$.

We find that the expressions for $\bar{n}$, ${\rm Var}\,n$, and $\nu_{\rm
eff}/\nu$ in the amplifying chaotic cavity differ from those in the dual
absorbing cavity by the substitution of $\gamma$ by $-\gamma$:
\begin{eqnarray}
&&\bar{n}=-\frac{\nu\alpha f\gamma}{1-\gamma},\label{barncavity2}\\
&&{\rm Var}\,n=\bar{n}+\nu(\alpha
f)^{2}\gamma^{2}\frac{\gamma^{2}-2\gamma+2}{(1-\gamma)^{4}},
\label{Varncavity2}\\
&&\frac{\nu_{\rm eff}}{\nu}=\frac{(1-\gamma)^{2}}{\gamma^{2}-2\gamma+2}.
\label{nueffcavity2}
\end{eqnarray}
The Bose-Einstein function $f$ is now to be evaluated at a negative
temperature, so that $f<0$. In Fig.\ \ref{nueff} we compare $\nu_{\rm eff}/\nu$
for amplifying and absorbing cavities. In the limit $\gamma\rightarrow 0$ the
two results coincide, but the $\gamma$-dependence is strikingly different:
While the ratio $\nu_{\rm eff}/\nu$ increases with $\gamma$ in the case of
absorption, it decreases in the case of amplification --- vanishing at the
laser threshold. Of course, close to the laser threshold [when $\gamma\gtequiv
1-(\Omega_{\rm c}\tau_{\rm dwell})^{-1/2}$] the approximation of a linear
amplifier breaks down and a non-linear treatment (along the lines of Ref.\
\cite{Bru69}) is required.

\begin{figure}
\hspace*{\fill}
\psfig{figure=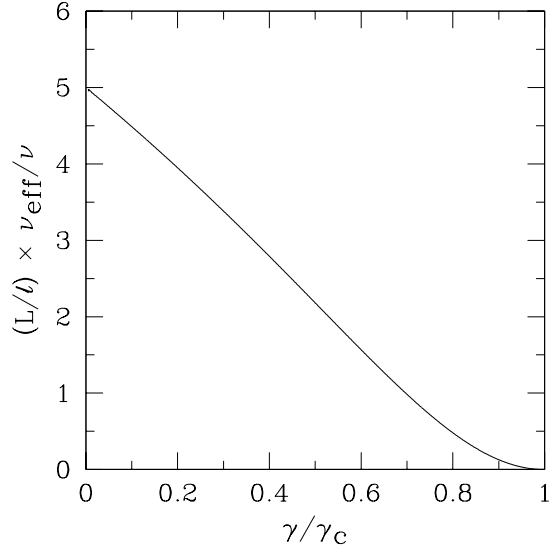,width=8cm}
\hspace*{\fill}
\caption[]{
Effective number of degrees of freedom for an amplifying disordered wave\-guide
of finite length $L$ (much greater than the transport mean free path $l$),
computed from Eq.\ (\protect\ref{nueffslab2}). The laser threshold occurs at
$\gamma=\gamma_{\rm c}\equiv (4\pi l/3L)^{2}$.
\label{figslab}
}
\end{figure}

For the amplifying disordered waveguide we can not use the infinite-length
formulas of Sec.\ \ref{waveguide}, because then we would be above threshold for
arbitrarily small $\gamma$. The counting distribution $P(n)$ at a finite $L$
requires the density of eigenvalues of the matrix $r\cdot r^{\dagger}+t\cdot
t^{\dagger}$, as explained in Sec.\ \ref{waveguidegeometry}. The density itself
is not known, but its first few moments have been calculated recently by
Brouwer \cite{Bro98}. This is sufficient to compute $\nu_{\rm eff}$, since we
only need the first two moments of $P(n)$. The results for $\gamma,\gamma_{\rm
c}\ll 1$ are\footnote{
Ref.\ \protect\cite{Bro98} considers an absorbing waveguide. The amplifying
case follows by changing the sign of the parameter $\gamma$. Eq.\ (13c) in
Ref.\ \protect\cite{Bro98} contains a misprint: The second and third term
between brackets should have, respectively, signs minus and plus instead of
plus and minus.}
\begin{eqnarray}
&&\bar{n}=-\nu\alpha f\frac{\sqrt{\gamma}}{\sin s}(1-\cos
s),\label{barnslab2}\\
&&{\rm Var}\,n=\bar{n}+\nu(\alpha f)^{2}\frac{\sqrt{\gamma}}{2\sin^{4}
s}(s-s\cos s+s\sin^{2} s+\sin s\nonumber\\
&&\hphantom{{\rm Var}\,n=}\mbox{}-3\sin^{3} s-\cos^{3} s\sin
s),\label{Varnslab2}\\
&&\frac{\nu_{\rm eff}}{\nu}=\frac{2\sqrt{\gamma}(1-\cos s)^{2}\sin^{2}
s}{s-s\cos s+s\sin^{2} s+\sin s-3\sin^{3} s-\cos^{3} s\sin
s},\label{nueffslab2}
\end{eqnarray}
where we have abbreviated $s=\pi\sqrt{\gamma/\gamma_{\rm c}}$. Fig.\
\ref{figslab} shows a plot of Eq.\ (\ref{nueffslab2}). Notice the limit
$\nu_{\rm eff}/\nu=5\,l/L$ for $\gamma/\gamma_{\rm c}\rightarrow 0$. The
reduction of the number of degrees of freedom on approaching the laser
threshold is qualitatively similar to that shown in Fig.\ \ref{nueff} for the
chaotic cavity.

\subsection{Broad-band detection}
\label{broadband}

In these applications we have assumed that only photons within a narrow
frequency interval $\delta\omega$ are detected. This simplifies the
calculations because the frequency dependence of the scattering matrix need not
be taken into account. In this subsection we consider the opposite extreme that
all frequencies are detected. We will see that this case of broad-band
detection is qualitatively similar to the case of narrow-band detection
considered so far.

We take a Lorentzian frequency dependence of the absorption or amplification
rate,
\begin{equation}
\gamma(\omega)=\frac{\gamma_{0}}{1+4(\omega-\omega_{0})^{2}/\Gamma^{2}}.
\label{gammaLorentz}
\end{equation}
The characteristic frequency $\Omega_{\rm c}$ for the scattering strengths is
defined by
\begin{equation}
\Omega_{\rm c}=\Gamma\sqrt{1+\gamma_{0}}.\label{OmegacGammarelation}
\end{equation}
The two frequencies $\Omega_{\rm c}$ and $\Gamma$ are essentially the same for
$\gamma_{0}\ltequiv 1$, but for $\gamma_{0}\gg 1$ the former is much bigger
than the latter. The reason is that what matters for the deviation of the
scattering strengths from zero is the relative magnitude of $\gamma(\omega)$
with respect to $1$, not with respect to $\gamma_{0}$. As in Sec.\
\ref{blackbody}, we assume that $\Omega_{\rm c}\ll\omega_{0}$, so that we may
neglect the frequency dependence of $N$ and $f$. The mean and variance of the
photo\-count in the long-time regime are given by
\begin{eqnarray}
&&\bar{n}=t\alpha f\int\frac{d\omega}{2\pi}\int
d\sigma\,\rho(\sigma,\omega)(1-\sigma), \label{barnbroad}\\
&&{\rm Var}\,n=\bar{n}+t(\alpha f)^{2}\int\frac{d\omega}{2\pi}\int
d\sigma\,\rho(\sigma,\omega)(1-\sigma)^{2}.\label{Varnbroad}
\end{eqnarray}
Again, we have assumed that $N$ is sufficiently large that sample-to-sample
fluctuations can be neglected and we may replace $\sum_{n}$ by $\int d\sigma$.
The scattering-strength density $\rho$ depends on $\omega$ through the rate
$\gamma(\omega)$.

In the absorbing, infinitely long disordered wave\-guide
$\rho\propto\sqrt{\gamma}$ for $\gamma\ll 1$, hence the integrands in Eqs.\
(\ref{barnbroad}) and (\ref{Varnbroad}) decay $\propto 1/|\omega-\omega_{0}|$
and the integrals over $\omega$ diverge. A cutoff is provided by the finite
length $L$ of the wave\-guide. When $\gamma$ drops below $(l/L)^{2}$, radiation
can be transmitted through the wave\-guide with little absorption. Only the
frequency range $|\omega-\omega_{0}|\ltequiv\Gamma(L/l)\sqrt{\gamma_{0}}$,
therefore, contributes effectively to the integrals (\ref{barnbroad}) and
(\ref{Varnbroad}). To leading order in $(L/l)\sqrt{\gamma_{0}}$ we can take the
infinite-$L$ result for $\rho$ with the cutoff we mentioned. The result (for
$(l/L)^{2}\ll\gamma_{0}\ll 1$) is
\begin{eqnarray}
&&\bar{n}=\frac{Nt\Gamma}{2\pi}\alpha
f\sqrt{\gamma_{0}}\left[\ln\left(\frac{L}{l}\sqrt{\gamma_{0}}\right)+{\cal
O}(1)\right]. \label{barnbroadslab}\\
&&{\rm Var}\,n=\bar{n}+\frac{Nt\Gamma}{2\pi}(\alpha
f)^{2}{\textstyle\frac{1}{2}}\sqrt{\gamma_{0}}
\left[\ln\left(\frac{L}{l}\sqrt{\gamma_{0}}\right)+{\cal O}(1)\right]. 
\label{Varnbroadslab}
\end{eqnarray}
If we write ${\rm Var}\,n=\bar{n}(1+\bar{n}/\nu_{\rm eff})$, as before, then
\begin{equation}
\frac{\nu_{\rm
eff}}{Nt\Gamma}=\pi^{-1}\sqrt{\gamma_{0}}
\left[\ln\left(\frac{L}{l}\sqrt{\gamma_{0}}\right)+{\cal O}(1)\right]. 
\label{nueffbroadslab}
\end{equation}
In the case of narrow-band detection considered in Sec.\ \ref{waveguide} we had
$\nu_{\rm eff}/Nt\delta\omega=\pi^{-1}\sqrt{\gamma_{0}}$ for
$(l/L)^{2}\ll\gamma_{0}\ll 1$. The difference with Eq.\ (\ref{nueffbroadslab})
(apart from the replacement of $\delta\omega$ by $\Gamma\approx\Omega_{\rm c}$)
is the logarithmic enhancement factor, but still $\nu_{\rm eff}\ll Nt\Gamma$
for $\gamma_{0}\ll 1$.

\begin{figure}
\hspace*{\fill}
\psfig{figure=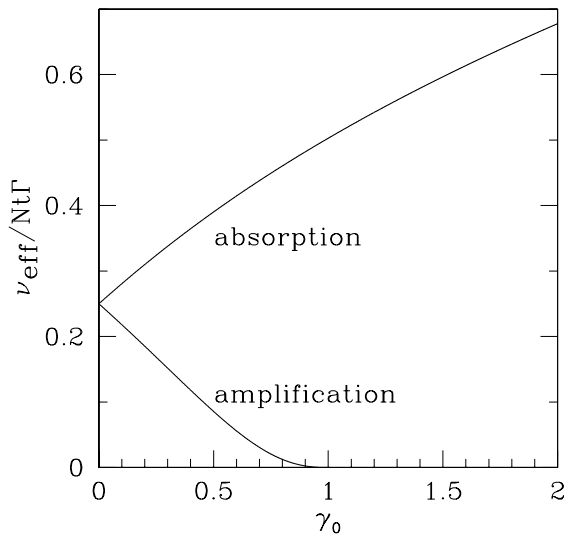,width=8cm}
\hspace*{\fill}
\caption[]{
Plot of Eq.\ (\protect\ref{nueffbroadcavity}) for the effective number of
degrees of freedom of an absorbing or amplifying chaotic cavity, in the case of
broad-band detection with a Lorentzian frequency dependence
$\gamma(\omega)=\gamma_{0} [1+4(\omega-\omega_{0})^{2}/\Gamma^{2}]^{-1}$ of the
absorption or amplification rate. For $\gamma_{0}\ltequiv 1$ the curves are
qualitatively similar those plotted in Fig.\ \protect\ref{nueff} for the case
of narrow-band detection. In the absorbing cavity $\nu_{\rm eff}$ increases
$\propto\sqrt{\gamma_{0}}$ as $\gamma_{0}\rightarrow\infty$, instead of
saturating as in Fig.\ \ref{nueff}, because the characteristic frequency
$\Omega_{\rm c}$ increases $\propto\sqrt{\gamma_{0}}$ in that limit [cf.\ Eq.\
(\protect\ref{OmegacGammarelation})].
\label{fignueffbroad}
}
\end{figure}

For the chaotic cavity, we can compute $\bar{n}$ and ${\rm Var}\,n$ directly
from the narrow-band results (\ref{barncavity}), (\ref{Varncavity}),
(\ref{barncavity2}), and (\ref{Varncavity2}), by substituting Eq.\
(\ref{gammaLorentz}) for $\gamma$ and integrating over $\omega$. There are no
convergence problems in this case. The results are
\begin{eqnarray}
&&\bar{n}=\pm Nt\Gamma\alpha f
\frac{\gamma_{0}}{4\sqrt{1\pm\gamma_{0}}},\label{barnbroadcavity}\\
&&{\rm Var}\,n=\bar{n}+Nt\Gamma(\alpha f)^{2}\gamma_{0}^{2}
\frac{9\gamma_{0}^{2}\pm 20\gamma_{0}+16}{64(1\pm\gamma_{0})^{7/2}},
\label{Varnbroadcavity}\\
&&\frac{\nu_{\rm
eff}}{Nt\Gamma}=\frac{4(1\pm\gamma_{0})^{5/2}}{9\gamma_{0}^{2}\pm
20\gamma_{0}+16}. \label{nueffbroadcavity}
\end{eqnarray}
The $\pm$ indicates that the plus sign should be taken for absorption and the
minus sign for amplification. The function (\ref{nueffbroadcavity}) is plotted
in Fig.\ \ref{fignueffbroad}. In the strongly absorbing limit
$\gamma_{0}\rightarrow\infty$, the effective number of degrees of freedom
$\nu_{\rm eff}\rightarrow\frac{4}{9}Nt\Omega_{\rm c}$, which up to a numerical
coefficient corresponds to the narrow-band limit $\nu_{\rm eff}\rightarrow
Nt\delta\omega/2\pi$ upon replacement of $\delta\omega$ by $\Omega_{\rm c}$.
The limit $\gamma_{0}\rightarrow 0$ is the same for absorption and
amplification, $\nu_{\rm eff}\rightarrow\frac{1}{4}Nt\Omega_{\rm c}$, again
corresponding to the narrow-band result $\nu_{\rm eff}\rightarrow
Nt\delta\omega/4\pi$ up to a numerical coefficient. Finally, the ratio
$\nu_{\rm eff}/Nt\Gamma$ tends to zero upon approaching the laser threshold
$\gamma_{0}\rightarrow 1$ in an amplifying system, similarly to the narrow-band
case. The qualitative behavior of $\nu_{\rm eff}/\nu$ is therefore the same for
broad-band and narrow-band detection.

\section{Conclusion}
\label{conclusion}

\subsection{Summary}
\label{summary}

In conclusion, we have shown that the photo\-detection statistics contains
substantially more information on the scattering properties of a medium than
its absorptivity. The mean photo\-count $\bar{n}$ is determined just by the
absorptivity, as dictated by Kirchhoff's law. Higher order moments of the
counting distribution, however, contain information on higher spectral moments
of the scattering strengths $\sigma_{n}$ (being the $N$ eigenvalues of the
scattering matrix product $S\cdot S^{\dagger}$). These higher moments are
independent of the absorptivity, which is determined by the first moment. While
the absorptivity follows from the radiative transfer equation, higher spectral
moments are outside of the range of that approach. We have used random-matrix
theory for their evaluation.

To measure these higher spectral moments, it is necessary that the counting
time $t$ is greater than the coherence time $1/\Omega_{\rm c}$ of the thermal
radiation (being the inverse of the absorption or amplification line width). It
is also necessary that the area of the photo\-cathode is greater than the
coherence area (being the area corresponding to one mode emitted by the
medium). Single-mode detection yields solely information on the absorptivity.
Multi-mode detection is unusual in photo\-detection experiments, but required
if one wants to go beyond Kirchhoff's law.

We have shown that the variance ${\rm Var}\,n$ of the photo\-count contains
information on the width of the density $\rho(\sigma)$ of scattering strengths.
We have computed this density for a disordered wave\-guide and for a chaotic
cavity, and find that it is very wide and strongly non-Gaussian. In an
absorbing medium, the deviations from Poisson statistics of independent
photo\-counts are small because the Bose-Einstein function is $\ll 1$ for all
practical frequencies and temperatures. Since the Poisson distribution contains
the mean photo\-count as the only parameter, one needs to be able to measure
the super-Poissonian fluctuations in order to obtain information beyond the
absorptivity. The deviations from Poisson statistics are easier to detect in an
amplifying medium, where the role of the Bose-Einstein function is played by
the relative population inversion of the atomic states.

We have shown that the super-Poissonian fluctuations in a linearly amplifying
random medium are much greater than would be expected from the mean
photo\-count. If we write ${\rm Var}\,n-\bar{n}=\bar{n}^{2}/\nu_{\rm eff}$,
then $\nu_{\rm eff}$ would equal $Nt\delta\omega/2\pi\equiv\nu$ if all $N$
modes reaching the photo\-detector would have the same scattering strength. (We
assume for simplicity that only a narrow band $\delta\omega$ is detected, in a
time $t$; For broad-band detection $\delta\omega$ should be replaced by
$\Omega_{\rm c}$.) The effective number $\nu_{\rm eff}$ of degrees of freedom
is much smaller than $\nu$ for a broad $\rho(\sigma)$, hence the anomalously
large fluctuations. On approaching the laser threshold, the ratio $\nu_{\rm
eff}/\nu$ goes to zero. In a conventional laser the noise itself increases with
increasing amplification rate because $\bar{n}$ increases, but $\nu_{\rm eff}$
does not change below the laser threshold. Typically, light is emitted in a
single mode, hence $\nu_{\rm eff}$ equals $t\delta\omega/2\pi$ independent of
the amplification rate. In a random laser a large number $N$ of cavity modes
contribute to the radiation, no matter how small the frequency window
$\delta\omega$, because the cavity modes overlap. The overlap is the
consequence of the much weaker confinement created by disorder in comparison to
that created by a mirror. The reduction of the number of degrees of freedom is
a quantum optical effect of overlapping cavity modes that should be observable
experimentally.

\subsection{Relation to Thouless number}
\label{Thouless}

The Thouless number $N_{\rm T}$ plays a central role in mesoscopic physics
\cite{Imr96}. It is a dimensionless measure of the coupling strength of a
closed system to the outside world,
\begin{equation}
N_{\rm T}\simeq\frac{1}{\tau_{\rm dwell}\Delta\omega}.\label{NTdef}
\end{equation}
(We use $\simeq$ instead of $=$ because we are ignoring numerical coefficients
of order unity.) As before, $\Delta\omega$ is the spacing of the
eigenfrequencies of the closed system and $\tau_{\rm dwell}$ is the mean time a
particle (electron or photon) entering the system stays inside. In a conducting
metal, $N_{\rm T}$ is the conductance in units of the conductance quantum
$e^{2}/h$. The metal-insulator transition occurs when $N_{\rm T}$ becomes of
order unity. It is assumed that there is no absorption or amplification, as is
appropriate for electrons.

For the two types of systems considered in this work, one has $N_{\rm T}\simeq
Nl/L$ for the disordered wave\-guide and $N_{\rm T}\simeq N$ for the chaotic
cavity. We notice that $N_{\rm T}$ is related to the effective number of
degrees of freedom in the limit of zero absorption and amplification,
\begin{equation}
\lim_{\gamma\rightarrow 0}\frac{\nu_{\rm eff}}{\nu}\simeq\frac{N_{\rm
T}}{N}.\label{nueffNTrelation}
\end{equation}
The ratio of $\nu_{\rm eff}$ to the black-body value $\nu$ is the same as that
of $N_{\rm T}$ to the number of propagating modes $N$. We believe that the
relation (\ref{nueffNTrelation}) holds for all random media, not just for those
considered here.

\section*{Acknowledgments}
I have benefitted from discussions with A. Lagendijk, R. Loudon, M. Patra, D.
S. Wiersma, and J. P. Woerdman. This research was supported by the
``Ne\-der\-land\-se or\-ga\-ni\-sa\-tie voor We\-ten\-schap\-pe\-lijk
On\-der\-zoek'' (NWO) and by the ``Stich\-ting voor Fun\-da\-men\-teel
On\-der\-zoek der Ma\-te\-rie'' (FOM).

\section*{Appendix. Scattering-strength density of a chaotic cavity}
\label{appendixrhocavity}

As derived in Ref.\ \cite{Bro97}, absorption in a chaotic cavity (with rate
$1/\tau_{\rm a}$) is statistically equivalent to the loss induced by a
fictitious wave\-guide that is weakly coupled to the cavity. The coupling has
transmission probability $\Gamma'$ for each of the $N'$ modes in the fictitious
wave\-guide. The equivalence requires the limit $N'\rightarrow\infty$,
$\Gamma'\rightarrow 0$, at fixed $N'\Gamma'=2\pi/\tau_{\rm a}\Delta\omega$
(with $\Delta\omega$ the spacing of the cavity modes).

\begin{figure}
\hspace*{\fill}
\psfig{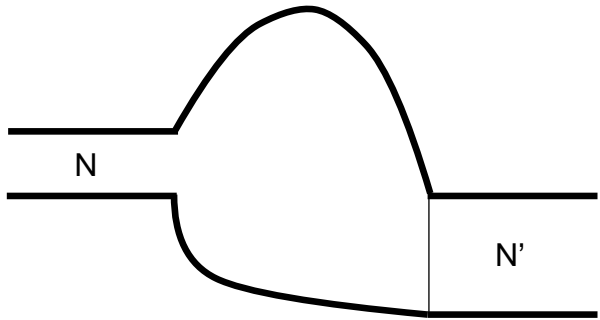}
\hspace*{\fill}
\smallskip\\
\caption[]{
An absorbing cavity with one opening is statistically equivalent to the
non-absorbing cavity with two openings shown here. The thin line in the second
opening indicates a barrier with transmission probability $\Gamma'$ for each of
the $N'$ modes in the waveguide attached to the opening. (The limit
$\Gamma'\rightarrow 0$, $N'\rightarrow\infty$ at fixed $N'\Gamma'$ is required
for the equivalence with absorption.)
\label{twoleadcavity}
}
\end{figure}

We are therefore led to consider the system illustrated in Fig.
\ref{twoleadcavity}: A chaotic cavity without absorption containing two
openings. One opening is coupled to an $N$-mode waveguide with transmission
probability $1$ per mode, the other opening is coupled to the fictitious
$N'$-mode wave\-guide with transmission probability $\Gamma'$ per mode. The
scattering strength $\sigma_{n}$ equals $1-T_{n}$, with $T_{n}$ an eigenvalue
of the transmission-matrix product $t\cdot t^{\dagger}$ ($t$ being the $N\times
N'$ transmission matrix from one wave\-guide to the other). The density of
transmission eigenvalues $\rho(T)=\langle\sum_{n}\delta(T-T_{n})\rangle$ can be
calculated in the large-$N$ regime using the perturbation theory of Ref.\
\cite{Bro96}. The scattering-strength density $\rho(\sigma)$ then follows from
$\sigma=1-T$.

The result is non-zero for $\sigma_{\rm min}<\sigma<\sigma_{\rm max}$, with the
definitions
\begin{eqnarray}
&&\sigma_{\rm min}=\left\{\begin{array}{cl}
\sigma_{-}&\;{\rm if}\;\;\gamma<1,\\
0&\;{\rm if}\;\;\gamma>1,
\end{array}\right.\label{sigmamaxdef}\\
&&\sigma_{\rm max}=\sigma_{+},\label{sigmamindef}\\
&&\sigma_{\pm}=\frac{8+20\gamma^{2}-\gamma^{4}\pm\gamma(8+\gamma^{2})^{3/2}}
{8(1+\gamma)^{3}}.\label{sigmapmdef}
\end{eqnarray}
[We use the same dimensionless absorption rate
$\gamma=N'\Gamma'/N=2\pi/N\tau_{\rm a}\Delta\omega$ as in Eq.\
(\ref{gammadef2}).] Inside this interval the scattering-strength density is
given by
\begin{eqnarray}
\rho(\sigma)&=&\frac{6N\sqrt{3}}{\pi}\biggl((a+b)^{1/3}-(a-b)^{1/3}\biggr)
\biggl(\biggl[(a+b)^{1/3}+(a-b)^{1/3}\nonumber\\
&&\mbox{}-2\gamma+2-6\sigma\biggr]^{2}+3\biggl[(a+b)^{1/3}-
(a-b)^{1/3}\biggr]^{2}\biggr)^{-1}, \label{fullresult}\\
a&=&(\gamma-1)^{3}+9(1+{\textstyle\frac{1}{2}}\gamma^{2})\sigma,\label{adef}\\
b&=&(3+3\gamma)^{3/2}[\sigma(\sigma-\sigma_{-})(\sigma_{+}-\sigma)]^{1/2}.
\label{bdef}
\end{eqnarray}
Eq.\ (\ref{fullresult}) is plotted in Fig.\ \ref{rhoplot} for several values of
$\gamma$.

\end{document}